\documentclass[envcountsame,final]{svjour3}
\usepackage{url}
\newcommand{\ket}[1]{|#1\rangle}
\newcommand{\bra}[1]{\langle #1 |}

\usepackage[T1]{fontenc}
\usepackage{txfonts}

\begin{document}
\journalname{Quantum Information Processing}
\title{Unitary Reconstruction of Secret for Stabilizer Based Quantum Secret Sharing}
\author{Ryutaroh Matsumoto}
\date{6 July 2017}
\institute{Ryutaroh Matsumoto \at Department of Information and Communication Engineering, Nagoya University, 464-8603 Japan\\
              ORCID: 0000-0002-5085-8879 \\
              \email{ryutaroh@rmatsumoto.org}\\
To be published in Quantum Information Processing. The final publication is available at Springer via \url{http://dx.doi.org/10.1007/s11128-017-1656-1}}
\maketitle
\begin{abstract}
  We propose a unitary procedure to reconstruct
  quantum secret for a quantum secret sharing scheme
  constructed from stabilizer quantum error-correcting
  codes. Erasure correcting procedures for stabilizer codes
  need to add missing shares for reconstruction
  of quantum secret while
  unitary reconstruction procedures for certain class of
  quantum secret sharing are known to
  work without adding missing shares.
  The proposed procedure also works without adding
  missing shares.
\end{abstract}

\section{Introduction}
Secret sharing (SS) \cite{shamir79} is a cryptographic scheme to
encode a secret to multiple shares being distributed to
participants, so that only qualified sets of participants
can reconstruct the original secret from their shares.
Traditionally both secret and shares were classical information
(bits). Several authors \cite{cleve99,gottesman00,hillery99,smith00}
extended the traditional SS to quantum one
so that a quantum secret can be encoded to quantum shares.

There was a difference between early pionieering works
\cite{cleve99,gottesman00,hillery99,smith00} of quantum SS.
The first quantum SS \cite{hillery99} was based on the controlled
teleportation \cite{PhysRevA.58.4394,PhysRevA.70.022329},
whose reconstruction of quantum secret involved
classical communication among participants.
On the other hand,
the others works \cite{cleve99,gottesman00,smith00} related
reconstruction to quantum error correction \cite{calderbank96,steane96},
and their reconstruction procedures were generally unitary
operations on quantum shares.
This paper studies reconstruction in the second category.

When we require unqualified sets of participants
to have zero information of the secret,
the size of each share must be larger than or equal to
that of secret.
By tolerating partial information leakage to
unqualified sets, the size of shares can be smaller
than that of secret. Such SS is called ramp SS
\cite{blakley85,yamamoto86}. The quantum ramp SS was
proposed by Ogawa et al.\ \cite{ogawa05}.
If an unqualified set has absolutely no information
about quantum secret (see \cite{ogawa05} for a formal
definition), it is called a forbidden set.

When we have a quantum error-correcting code (QECC)
of length $n$, use it for quantum secret sharing and
it can correct erasures in
a set $\overline{J} \subset \{1$, \ldots, $n\}$,
it was shown \cite{cleve99,gottesman00} that
$J = \{1$, \ldots, $n\} \setminus \overline{J}$
is a qualified set and $\overline{J}$ is a forbidden set.
The above statement also holds for quantum ramp SS \cite{ogawa05}.
In such a situation,
a straightforward method for the set $J$ of participants
to reconstruct quantum secret is as follows:
Firstly, initialize quantum systems in
$\overline{J}$ to any quantum states and
apply the erasure decoding procedure of QECC.
This method is wasteful because decoding procedures usually
involve measurement and they also need to attach $|\overline{J}|$
extra quantum systems.
For example, if $|J| = 70$ and $|\overline{J}|=30$,
adding 30 quantum systems and performing measurement on
100 systems are wasteful.

To overcome this waste, unitary reconstruction methods
were proposed for previous proposals of quantum SS
\cite{cleve99,matsumoto14qss,ogawa05,matsumoto14strong}.
On the other hand, while quantum SS constructed from
the stabilizer QECC had been already studied \cite{marin13,markham08,sarvepalli12},
no unitary reconstruction procedure has been
proposed for stabilizer based quantum SS.
Stabilizer based quantum SS is important
because it can realize access structures that cannot
be realized by quantum SS based on CSS codes \cite{calderbank96,steane96}.
For example, only the $[[5,1,3]]$ binary stabilizer QECC
can realize quantum SS distributing
1 qubit of secret to 5 participants receiving
1-qubit shares and allowing only
3 or more participants to reconstruct secret.
In addition, when sharing classical secret,
it was recently shown that stabilizer QECC can realize an access structure
that cannot be realized by classical information processing
\cite{matsumoto17impossible}.

In this paper, we propose a unitary reconstruction
method that can be executed by a qualified set
$J$ of participants without adding extra quantum systems.
In Section 2, we introduce notations of stabilizer QECC
and prove some properties of stabilizer QECC used later
in the proposed reconstruction procedure.
Section 3 describes the proposed procedure.
Section 4 gives an explicit computational example
of the proposed procedure applied to
the well-known $[[5,1,3]]$ binary stabilizer QECC.
In the Appendix, we discuss the security of quantum SS
based on stabilizer QECCs.

\section{Preliminaries}
\subsection{Notations for Stabilizer Codes}
Let $q$ be a prime power,
and we consider the $q$-dimensional
complex linear space $\mathbf{C}_q$.
A quantum system whose state is expressed by $\mathbf{C}_q$
is called a qudit in this paper.
Each share is assumed to be a qudit,
and quantum secret consists of one or more qudits.
If quantum secret has two or more qudits,
the quantum SS becomes a ramp scheme.
We fix a $q$-ary stabilizer QECC encoding
$k$ qudits to $n$ qudits.
The materials in this subsection is not new at all,
and can be found in, for example, \cite{ashikhmin00,grassl11}.
Its stabilizer can be expressed as
an $(n-k)$-dimensional $\mathbf{F}_q$-linear
subspace $C$ of $\mathbf{F}_q^{2n}$,
where $\mathbf{F}_q$ is the finite field with $q$ elements.

For two vectors $\vec{x} = (a_1$, $b_1$, \ldots, $a_n$, $b_n)$
and $\vec{y} = (a'_1$, $b'_1$, \ldots, $a'_n$, $b'_n) \in \mathbf{F}_q^{2n}$,
we define its symplectic inner product as
\begin{equation}
  \langle \vec{x}, \vec{y} \rangle =
  \sum_{i=1}^n a_i b'_i - a'_ib_i. \label{eq1}
\end{equation}
Let $C^\perp = \{ \vec{x} \in \mathbf{F}_q^{2n}
\mid \forall \vec{y} \in C$, $\langle \vec{x}$, $\vec{y}\rangle=0\}$.
Then we have $C^\perp \supset C$ and $\dim C^\perp = n+k$.

\subsection{Qualified Sets and Related Properties}
To use any reconstruction procedure,
the set $J$ of participants must be qualified to reconstruct the secret.
In this subsection,
we clarify a necessary and sufficient condition for
qualified sets and related properties that are later used for
the proposed reconstruction procedure.

For a set $J \subset \{1$, \ldots, $n\}$ of
participants to be qualified,
the erasures in $\overline{J}$ must be decodable,
where an erasure means a quantum error with known location.
In other words, when the errors are only in $\overline{J}$,
the stabilizer QECC defined by the stabilizer
$C \subset \mathbf{F}_q^{2n}$ must be able to
correct the error.

Let $\vec{g}_1$, \ldots, $\vec{g}_{n-k}$
be a basis of $C$.
A quantum error can also be identified with
a vector $\vec{e} =(a_1$, $b_1$, \ldots,
$a_n$, $b_n) \in \mathbf{F}_q^{2n}$
(see, for example, \cite{ashikhmin00,grassl11}).
Measurement in the standard decoding procedure
gives the symplectic inner products $\langle \vec{e}$,
$\vec{g}_i\rangle$ for $i=1$, \ldots, $n-k$.
Let $\mathbf{F}_q^{\overline{J}} =
\{ (a_1$, $b_1$, \ldots,
$a_n$, $b_n) \in \mathbf{F}_q^{2n} \mid $ $
j \in J \Rightarrow (a_j$, $b_j) = (0$, $0) \}$ and
$\mathbf{F}_q^{J} =
\{ (a_1$, $b_1$, \ldots,
$a_n$, $b_n) \in \mathbf{F}_q^{2n} \mid $ $
j \in \overline{J} \Rightarrow (a_j$, $b_j) = (0$, $0) \}$.
Observe that $\dim \mathbf{F}_q^J = 2|J|$ and $\dim \mathbf{F}_q^{\overline{J}} = 2|\overline{J}|$.

Under the assumption $j \in J \Rightarrow (a_j$, $b_j) = (0,0)$ for
$\vec{e}$,
we can correct all errors $\vec{e} \in \mathbf{F}_q^{\overline{J}}$
if and only if the implication
\begin{equation}
  \forall i, \langle \vec{e}, \vec{g}_i\rangle = 0
  \Rightarrow \vec{e} \in C \label{eq2}
\end{equation}
holds.
The condition (\ref{eq2}) implies
 (with the assumption that
errors belong to $\overline{J}$)
\[
C^\perp \cap \mathbf{F}_q^{\overline{J}} \subseteq C \cap \mathbf{F}_q^{\overline{J}}.
\]
On the other hand, the assumption
$C^\perp \supset C$ implies  
\[
C^\perp \cap \mathbf{F}_q^{\overline{J}} \supseteq C \cap \mathbf{F}_q^{\overline{J}}.
\]
Therefore the condition (\ref{eq2}) is equivalent to
\begin{equation}
  C^\perp \cap \mathbf{F}_q^{\overline{J}} = C \cap \mathbf{F}_q^{\overline{J}}.
  \label{eq3}
\end{equation}

We will study the linear spaces
consisting of qudits in $J$ or $\overline{J}$ of quantum codewords.
Let $Q(C)\subset \mathbf{C}_q^{\otimes n}$, $Q(C\cap \mathbf{F}_q^{\overline{J}})
\subset \mathbf{C}_q^{\otimes |\overline{J}|}$,
$Q(C\cap \mathbf{F}_q^J)\subset \mathbf{C}_q^{\otimes |J|}$
be stabilizer QECCs defined by $C$, $C\cap \mathbf{F}_q^{\overline{J}}$,
and $C\cap \mathbf{F}_q^J$, respectively.
When we consider qudits in $J$ (resp.\ $\overline{J}$) of codewords in $Q(C)$,
their quantum states are  density matrices whose row spaces
are contained in $Q(C\cap \mathbf{F}_q^J)$ (resp.\ $Q(C\cap \mathbf{F}_q^{\overline{J}})$).

In order to evaluate their dimensions, firstly
we have to evaluate $\dim C\cap \mathbf{F}_q^J$ and $\dim
C\cap \mathbf{F}_q^{\overline{J}}$,
where $\dim C\cap \mathbf{F}_q^J$ denotes the dimension
of the linear space $C\cap \mathbf{F}_q^J$.
We have
\begin{eqnarray}
  |J|-k - |\overline{J}| &\leq& \dim C\cap \mathbf{F}_q^J \label{eq5}\\
  & \leq & |J|-k. \label{eq6}
\end{eqnarray}

The linear space $C\cap \mathbf{F}_q^J$
consists of vectors in $C$ whose $(2j-1)$-th component and
$2j$-th component are zero if $j \in \overline{J}$,
which implies
$\dim C - \dim C\cap \mathbf{F}_q^J \leq 2 |\overline{J}|$.
Eq.\ (\ref{eq5}) holds because
\begin{eqnarray*}
  && \underbrace{\dim C}_{=n-k} - 2|\overline{J}| \leq \dim C\cap \mathbf{F}_q^J\\
  &\Leftrightarrow &  \underbrace{n-|\overline{J}|}_{=|J|} -k -
    |\overline{J}| \leq \dim C\cap \mathbf{F}_q^J\\
    &\Leftrightarrow &  |J|-k - |\overline{J}| \leq \dim C\cap \mathbf{F}_q^J.
\end{eqnarray*}

For $\vec{x} = (a_1$, $b_1$, \ldots, $a_n$, $b_n) \in \mathbf{F}_q^{2n}$,
let $P_{\overline{J}} (\vec{x}) = (a_j, b_j)_{j \in \overline{J}}$,
that is, the projection to the index set $\overline{J}$.
Then we have $C\cap \mathbf{F}_q^J = C \cap \ker(P_{\overline{J}})$
and $\dim C\cap \mathbf{F}_q^J + \dim P_{\overline{J}}(C)
= \dim C$,
which implies
\begin{equation}
  \dim P_{\overline{J}}(C) =  (n-k) - \dim C\cap \mathbf{F}_q^J. \label{eq100}
\end{equation}

Suppose that Eq.\ (\ref{eq6}) does not hold,
then we have $\dim P_{\overline{J}}(C) < |\overline{J}|$
by Eq.\ (\ref{eq100}) and the equality $n=|J|+|\overline{J}|$.
Since $C^\perp \cap \mathbf{F}_q^{\overline{J}} = P_{\overline{J}}(C)^\perp$
($\perp$ in $P_{\overline{J}}(C)^\perp$
is considered in $\mathbf{F}_q^{2|\overline{J}|}$),
we have $\dim C^\perp \cap \mathbf{F}_q^{\overline{J}}
= 2|\overline{J}| - \dim P_{\overline{J}}(C) > |\overline{J}|$.
The last inequality implies $\dim C^\perp \cap \mathbf{F}_q^{\overline{J}}
> |\overline{J}| > \dim P_{\overline{J}}(C) \geq \dim C \cap \mathbf{F}_q^{\overline{J}}$ because $P_{\overline{J}}(C) \supseteq C \cap \mathbf{F}_q^{\overline{J}}$.
The inequality $\dim C^\perp \cap \mathbf{F}_q^{\overline{J}}
> \dim C \cap \mathbf{F}_q^{\overline{J}}$ contradicts with Eq.\ (\ref{eq3}).
So we see that Eq.\ (\ref{eq6}) is true
when $J$ is a qualified set.

In light of Eqs.\ (\ref{eq5}) and (\ref{eq6}),
let $\dim C\cap \mathbf{F}_q^J = |J| - k - \ell$.
Then $Q(C\cap \mathbf{F}_q^J)$ encodes $k+\ell$ qudits
to $|J|$ qudits.

We consider $\dim
C\cap \mathbf{F}_q^{\overline{J}}$.
By Eq.\ (\ref{eq3}) we have
\begin{eqnarray*}
  && \dim
  C\cap \mathbf{F}_q^{\overline{J}} \\
  &=& \dim C^\perp\cap \mathbf{F}_q^{\overline{J}}\\
  &=& \dim C^\perp - \dim \underbrace{P_J(C^\perp)}_{=(C\cap \mathbf{F}_q^{J})^\perp
  \, \mathrm{ in }\, \mathbf{F}_q^{2|J|}}\\
  &=& \underbrace{\dim C^\perp}_{=n+k} - (2|J| - \dim C\cap \mathbf{F}_q^{J})\\
  &=& n+k - 2|J| + \underbrace{\dim C\cap \mathbf{F}_q^{J}}_{=|J|-(k+\ell)}\\
  &=& |\overline{J}| - \ell,
\end{eqnarray*}
which means that 
$Q(C\cap \mathbf{F}_q^{\overline{J}})$ encodes $\ell$ qudits
to $|\overline{J}|$ qudits.
Readers might wonder if $\ell=|\overline{J}|$ is always true.
The equality $\ell = |\overline{J}|$ usually holds as we will
see in Section \ref{sec4} with an example.
But $\ell = |\overline{J}|$ is sometimes false in general cases,
for example, consider an unpractical stabilizer QECC
whose codewords are always set to $\ket{00\cdots 0}$ in $\overline{J}$,
which gives $\ell=0$.

\section{Proposed Unitary Reconstruction}
For ease of presentation, without loss of generality
we may assume $\overline{J} = \{1$, \ldots, $|\overline{J}|\}$
and $J = \{ |\overline{J}| + 1$, \ldots, $n\}$,
by reordering indices.
Let
\begin{equation}
  \{ \ket{\vec{i^{(k)}}} \mid \vec{i^{(k)}} \in \mathbf{F}_q^k \}
  \label{eq10}
\end{equation}
be an orthonormal basis (ONB) of $\mathbf{C}_q^{\otimes k}$,
let $\ket{\psi(\vec{i^{(k)}})} \in Q(C)$ the quantum codeword
corresponding to $\ket{\vec{i^{(k)}}}$.
Let
\begin{equation}
  \{ \ket{\varphi_{\overline{J}}(\vec{i^{(\ell)}})}
  \mid \vec{i^{(\ell)}} \in \mathbf{F}_q^\ell\}
  \label{eq9}
\end{equation}
be an ONB of $Q(C\cap \mathbf{F}_q^{\overline{J}})$.
Then
\begin{equation}
(\ket{\varphi_{\overline{J}}(\vec{i^{(\ell)}})}
\bra{\varphi_{\overline{J}}(\vec{i^{(\ell)}})}\otimes I_J)
\ket{\psi(\vec{i^{(k)}})} \label{eq11}
\end{equation}
have the same nonzero length for all $\vec{i^{(k)}}$ and $\vec{i^{(\ell)}}$,
where $I_J$ is the identity matrix on qudits in $J$.
Because otherwise the Holevo information between
classical information $\vec{i^{(k)}}$ and the qudits in $\overline{J}$
would have strictly positive value which contradicts by \cite{ogawa05}
to
our assumption that $J$ is a qualified set

Define a state vector $\ket{\varphi_J(\vec{i^{(k)}}, \vec{i^{(\ell)}})}
\in Q(C\cap \mathbf{F}_q^J)$ by
\begin{equation}
\ket{\varphi_{\overline{J}}(\vec{i^{(\ell)}})}\ket{\varphi_J(\vec{i^{(k)}}, \vec{i^{(\ell)}})}
=
\frac{(\ket{\varphi_{\overline{J}}(\vec{i^{(\ell)}})}
\bra{\varphi_{\overline{J}}(\vec{i^{(\ell)}})}\otimes I_J)
\ket{\psi(\vec{i^{(k)}})}}{\|(\ket{\varphi_{\overline{J}}(\vec{i^{(\ell)}})}
\bra{\varphi_{\overline{J}}(\vec{i^{(\ell)}})}\otimes I_J)
\ket{\psi(\vec{i^{(k)}})}\|}. \label{eq8}
\end{equation}
Then $\ket{\varphi_J(\vec{i^{(k)}}, \vec{i^{(\ell)}})}$
is of length one and orthogonal to each other for
different $(\vec{i^{(k)}}$, $\vec{i^{(\ell)}})$.
Therefore
\begin{equation}
  \{
  \ket{\varphi_J(\vec{i^{(k)}}, \vec{i^{(\ell)}})}
  \mid \vec{i^{(k)}} \in \mathbf{F}_q^k,
  \vec{i^{(\ell)}} \in \mathbf{F}_q^\ell \} \label{eq12}
\end{equation}
is an ONB of $Q(C\cap \mathbf{F}_q^J)$.

By using the above notations we can express
\begin{equation}
  \ket{\psi(\vec{i^{(k)}})}
  =
  \frac{1}{\sqrt{q^\ell}}
    \sum_{\vec{i^{(\ell)}} \in \mathbf{F}_q^\ell}
    \ket{\varphi_{\overline{J}}(\vec{i^{(\ell)}})}
    \ket{\varphi_J(\vec{i^{(k)}}, \vec{i^{(\ell)}})}. \label{eq13}
\end{equation}
We can define a unitary operation $U_{\mathrm{rec}}$
from $Q(C\cap \mathbf{F}_q^J)$ to $Q(C\cap \mathbf{F}_q^{\overline{J}})\otimes
\mathbf{C}_q^{\otimes k}$,
sending
$\ket{\varphi_J(\vec{i^{(k)}}, \vec{i^{(\ell)}})}$
to $\ket{\varphi_{\overline{J}}(\vec{i^{(\ell)}})}
\ket{\vec{i^{(k)}}}$,
because both $\{
\ket{\varphi_J(\vec{i^{(k)}}, \vec{i^{(\ell)}})} \mid
\vec{i^{(k)}} \in \mathbf{F}_q^k$,
$\vec{i^{(\ell)}} \in \mathbf{F}_q^\ell \}$ and
$\{ \ket{\varphi_{\overline{J}}(\vec{i^{(\ell)}})}\ket{\vec{i^{(k)}}}
\mid \vec{i^{(k)}} \in \mathbf{F}_q^k$,
$\vec{i^{(\ell)}} \in \mathbf{F}_q^\ell \}$
are ONBs with the same number of quantum state vectors in them.

Suppose that quantum secret is
\[
\sum_{\vec{i^{(k)}} \in \mathbf{F}_q^k}
\alpha(\vec{i^{(k)}}) \ket{\vec{i^{(k)}}},
\]
where $\alpha(\vec{i^{(k)}})$ are complex coefficients.
Then the whole quantum state of all shares is, by Eq.\ (\ref{eq13}),
\[
\sum_{\vec{i^{(k)}} \in \mathbf{F}_q^k}\alpha(\vec{i^{(k)}})
\frac{1}{\sqrt{q^\ell}}
    \sum_{\vec{i^{(\ell)}} \in \mathbf{F}_q^\ell}
    \ket{\varphi_{\overline{J}}(\vec{i^{(\ell)}})}
    \ket{\varphi_J(\vec{i^{(k)}}, \vec{i^{(\ell)}})}.
\]
Applying $U_{\mathrm{rec}}$ on the qualified set $J$ yields
\begin{equation}
\left(\frac{1}{\sqrt{q^\ell}}
    \sum_{\vec{i^{(\ell)}} \in \mathbf{F}_q^\ell}
    \ket{\varphi_{\overline{J}}(\vec{i^{(\ell)}})}
    \ket{\varphi_{\overline{J}}(\vec{i^{(\ell)}})}\right)\otimes 
    \sum_{\vec{i^{(k)}} \in \mathbf{F}_q^k}\alpha(\vec{i^{(k)}})\ket{\vec{i^{(k)}}}.
    \label{eq:last}
\end{equation}
Equation (\ref{eq:last})
means that the quantum secret is reconstructed
in the rightmost $k$ qudits,
and that it is unentangled from the rest of qudits.

\section{Explicit Computational Example of
  the $[[5,1,3]]$ Binary Stabilizer QECC}\label{sec4}
Since our presentation of the proposed procedure is slightly abstract,
in this section we will see an explicit computational example with
the $[[5,1,3]]$ binary stabilizer QECC.
According to \cite{gottesmanthesis},
the $[[5,1,3]]$ binary stabilizer QECC
encodes $\ket{0}$ to
\begin{eqnarray*}
&& \ket{\psi(0)} \\
&= & \ket{00000} + \ket{10010} + \ket{01001} + \ket{10100}  \\
& & \mbox{} + \ket{01010} - \ket{11011} - \ket{00110} - \ket{11000} \\
& & \mbox{} - \ket{11101} - \ket{00011} - \ket{11110} - \ket{01111} \\
& & \mbox{} - \ket{10001} - \ket{01100} - \ket{10111} + \ket{00101}, 
\end{eqnarray*}
and $\ket{1}$ to
\begin{eqnarray*}
&& \ket{\psi(1)} \\
& = & \ket{11111} + \ket{01101} + \ket{10110} + \ket{01011}  \\
& & \mbox{} + \ket{10101} - \ket{00100} - \ket{11001} - \ket{00111}  \\
& & \mbox{} - \ket{00010} - \ket{11100} - \ket{00001} - \ket{10000} \\
& & \mbox{} - \ket{01110} - \ket{10011} - \ket{01000} + \ket{11010}.
\end{eqnarray*}
According to \cite[Table 3.2]{gottesmanthesis},
the corresponding stabilizer $C \subset \mathbf{F}_2^{10}$
is generated by
\begin{eqnarray*}
  \vec{g}_1 &=& (1,0,0,1,0,1,1,0,0,0),\\
  \vec{g}_2 &=& (0,0,1,0,0,1,0,1,1,0),\\
  \vec{g}_3 &=& (1,0,0,0,1,0,0,1,0,1),\\
  \vec{g}_4 &=& (0,1,1,0,0,0,1,0,0,1).
\end{eqnarray*}
Since it can correct any two erasures,
we can set $J=\{3$, $4$, $5\}$ and
$\overline{J} = \{1$, $2\}$.
Since $C \cap \mathbf{F}_2^{\overline{J}} = C^\perp \cap \mathbf{F}_2^{\overline{J}}$
are zero linear spaces, we can see that Eq.\ (\ref{eq3}) holds
and $\ell=2$.
We can choose $\ket{\varphi_{\overline{J}}(\vec{i^{(\ell)}})}$ of
Eq.\ (\ref{eq9})
as $\ket{\varphi_{\overline{J}}(00)} =\ket{00}$,
$\ket{\varphi_{\overline{J}}(01)} =\ket{01}$,
$\ket{\varphi_{\overline{J}}(10)} =\ket{10}$, and
$\ket{\varphi_{\overline{J}}(11)} =\ket{11}$.
Then $\ket{\varphi_J(\vec{i^{(k)}}, \vec{i^{(\ell)}})}$
of Eq.\ (\ref{eq8}) become the following states:
\begin{eqnarray*}
  \ket{\varphi_J(0,00)} &=& \frac{1}{2}(\ket{000}-\ket{110}-\ket{011}+\ket{101}),\\
  \ket{\varphi_J(0,01)} &=& \frac{1}{2}(\ket{001}+\ket{010}-\ket{111}-\ket{100}),\\
  \ket{\varphi_J(0,10)} &=& \frac{1}{2}(\ket{010}+\ket{100}-\ket{001}-\ket{111}),\\
  \ket{\varphi_J(0,11)} &=& \frac{1}{2}(-\ket{011}-\ket{000}-\ket{101}-\ket{110}),\\
  \ket{\varphi_J(1,00)} &=& \frac{1}{2}(-\ket{100}-\ket{111}-\ket{010}-\ket{001}),\\
  \ket{\varphi_J(1,01)} &=& \frac{1}{2}(\ket{101}+\ket{011}-\ket{110}-\ket{000}),\\
  \ket{\varphi_J(1,10)} &=& \frac{1}{2}(\ket{110}+\ket{101}-\ket{000}-\ket{011}),\\
  \ket{\varphi_J(1,11)} &=& \frac{1}{2}(\ket{111}-\ket{001}-\ket{100}+\ket{010}).
\end{eqnarray*}
The unitary reconstruction $U_{\mathrm{rec}}$ works as follows:
\[
U_{\mathrm{rec}}\ket{\varphi_J(i_1,i_2i_3)} = \ket{i_2 i_3}\ket{i_1}.
\]
If the quantum secret is $\alpha(0)\ket{0} + \alpha(1)\ket{1}$,
then the quantum state of all shares is $\alpha(0)\ket{\psi(0)} + \alpha(1)\ket{\psi(1)}$.
Application of $U_{\mathrm{rec}}$ to the 3rd, the 4th and the 5th qubits
of $\alpha(0)\ket{\psi(0)} + \alpha(1)\ket{\psi(1)}$ gives
\[
\frac{1}{2}(\ket{0000}+\ket{0101}+\ket{1010}+\ket{1111})(\alpha(0)\ket{0} + \alpha(1)\ket{1}),
\]
which means that the 3rd, the 4th and the 5th participants
successfully reconstructed the quantum secret
into the 5th qubit.
Also observe that after the reconstruction the 5th qubit
is completely unentangled from the rest of qubits.
Since the proposed procedure only interacts with
the 3rd to the 5th qubits,
even if there are errors in the 1st and the 2nd qubits,
after reconstruction we obtain
$\alpha(0)\ket{0} + \alpha(1)\ket{1}$ at the 5th qubit.

\appendix
\section{Security Analysis}
For the completeness of this paper,
in this appendix we discuss the security of
quantum SS based on stabilizer QECCs.
For the security analysis of quantum secret sharing
based on quantum error correction, such as \cite{cleve99,gottesman00},
we need to clarify (a) which share sets are qualified
(being able to reconstruct secret perfectly) and
(b) which share sets are forbidden (having no information about
secret). The characterization of qualified sets in the proposed scheme
is given by Eq.\ (\ref{eq3}).
Observe also that from a given basis
$\vec{g}_1$, \ldots, $\vec{g}_{n-k}$ of $C$,
we can easily verify by standard linear algebra
whether or not Eq.\ (\ref{eq3}) holds
for an arbitrarily given share set $J$.
The characterization of forbidden sets
also immediately follows from the fact that
a share set is forbidden if and only if the rest of
shares is qualified, as shown in \cite{cleve99,gottesman00,ogawa05}.

\begin{acknowledgement}
The author would like to thank reviewers' comments that
improved this paper significantly.
This research is partly supported by the JSPS Grant  
 No.\ 26289116.
\end{acknowledgement}


\end{document}